\documentclass{article}
\usepackage{float}
\usepackage[english]{babel}
\usepackage{natbib}
\usepackage{comment}

\usepackage[letterpaper,top=2cm,bottom=2cm,left=3cm,right=3cm,marginparwidth=1.75cm]{geometry}

\usepackage{amsmath}
\usepackage{graphicx}
\usepackage{subcaption}
\usepackage[colorlinks=true, allcolors=blue]{hyperref}
\usepackage{float}

\title{Quantile Regression Tree}
\author{Jaachinma Okafor, Lateefah Isegen, Ark Ifeanyi}

\begin{document}
\maketitle

\begin{abstract}
This study introduces and evaluates the Quantile Regressor Tree (QRT), a novel methodology merging the robust characteristics of quantile regression with the versatility of decision trees. The quantile regressor tree introduces non-linearity to the quantile regression due to the splitting by features in the decision tree, enhancing flexibility while maintaining interpretability. The quantile regression tree gives a parametric and non-parametric mixture of estimating conditional quantiles for high-dimensional predictor variables. 
\end{abstract}

\section{Introduction}
Quantile regression has emerged as a powerful tool in statistical modeling, offering a more comprehensive view of the conditional distribution of a response variable than the traditional mean regression \citep{koenker2001quantile}. This approach is particularly valuable in understanding the behavior of a dependent variable across different quantiles, rather than just focusing on the mean. It provides insights into the distribution tails, crucial in fields like finance, and predictive maintenance \citep{phm1,phm2} where extreme values can have significant impacts.

This paper introduces a novel approach to quantile regression: the quantile regressor tree (QRT). Combining the robustness of decision trees with the flexibility of quantile regression, this model aims to provide a more nuanced understanding of conditional quantile functions. The quantile regressor tree is designed to capture complex, non-linear relationships in the data, making it a versatile tool for various predictive modeling tasks.

Our implementation builds upon the foundations of decision tree algorithms, extending their functionality to accommodate quantile regression. By integrating the quantile regressor from the sci-kit-learn library with the decision tree model algorithm, we have developed a model that can adaptively partition the feature space and provide quantile estimates at each terminal node. This approach allows for capturing different behaviors at various sections of the response variable's distribution, making it an excellent choice for datasets with heterogeneous variances or non-normal error distributions.

The paper presents a comprehensive evaluation of the quantile regressor tree, benchmarking its performance against the traditional quantile regression model, and also comparing the predictive accuracy of the conditional median values of the QRT against the predictive accuracy of linear regression and random forest regressors conditional mean values. Through empirical analysis, we demonstrate the efficacy of our model in various scenarios, emphasizing its utility in capturing conditional quantile functions effectively.

\section{Literature Review}
Quantile regression, first introduced by 
\cite{koenker1978regression}, has evolved significantly, offering a more detailed view of possible causal relationships in data. Traditional regression methods, like Ordinary Least Squares (OLS) \citep{craven2011ordinary}, focus on estimating the mean of the dependent variable. In contrast, quantile regression enables the estimation of conditional quantile functions, thereby providing a comprehensive understanding of the distribution of the response variable. This approach is especially beneficial for datasets with non-normal residuals or heteroscedasticity. Studies by \cite{koenker2001quantile}
have furthered the application of quantile regression in various fields, highlighting its flexibility and robustness in different contexts.

Decision trees, a form of non-parametric supervised learning, are widely used for classification and regression tasks \citep{rokach2005decision}. The allure of decision trees lies in their simplicity and interpretability. Regression trees, a variant of decision trees, are used for predicting continuous outcomes. 
\citep{breiman1984cart} introduced the CART (Classification and Regression Trees) algorithm, which laid the groundwork for numerous tree-based methods. Subsequent advancements, like random forests by \cite{breiman2001random},
have improved the predictive accuracy and generalizability of tree-based models.

The integration of quantile regression into decision trees is a relatively recent development. 
\cite{meinshausen2006quantile} proposed quantile regression forests, which extended the concept of Random Forests to quantile regression. This approach aimed to combine the predictive power of tree ensembles with the ability to estimate conditional quantiles. The difference between \cite{meinshausen2006quantile}'s approach and our approach is in the estimation of the model. In contrast to their approach that uses the conditional distribution function to estimate the leaf nodes, our model makes use of quantile regression to estimate the leaf nodes. Therefore our approach is a mixture of parametric and non-parametric models. Also, we build on decision trees and not random forest models.

\section{Model}
Let $D = \{(x_1, y_1), (x_2, y_2), \ldots, (x_n, y_n)\}$
be independent observations of a dataset. Where n represents the total number of samples, $x_i$ represents a feature vector for the i-th row, $x_i \in R^d$, where d denotes the dimensionality of the feature space and $y_i$ is the i-th observation of the continuous dependent variable.

For linear regression, we have:
\begin{equation}\label{eqn:1}
    y_i = \beta_0 + x_i^1 \beta_1 + e
\end{equation}
where
\begin{equation}\label{eqn:2}
    [\beta_0 \beta_1] = \arg\min_{\beta} \frac{1}{n} \sum_{i=1}^{n} L(y_i, f(x_i;\beta))
\end{equation}\label{eqn:3}
and 
\begin{equation}
    L(y_i, f(x_i;\beta)) = (y_i - \beta_0 - x_i^1 \beta_1)^2
\end{equation}

For quantile regression, we have the same equation as Equation(\ref{eqn:1}) and Equation(\ref{eqn:2}) but Equation(\ref{eqn:3}) becomes:
\begin{equation}
    L(y_i, f(x_i;\beta)) = \rho_r(y_i - \beta_0 - x_i^1 \beta_1)
\end{equation} 
where r is the quantile of estimation and $\rho_r(*)$ is a loss function as specified in \cite{koenker1978regression}

For a decision regression tree, a tree is grown using selection criterion as described by \cite{rokach2005decision} where the prediction of a new data point is the expected conditional mean of the final leaf node in which the data point falls.

\subsection{Quantile Regression Tree Model}
The quantile regression tree like a decision regression tree performs regression by recursively partitioning the dataset based on its  features.
\subsubsection{Split Criterion}
Given $x_i \in R^d$, a quantile regression tree performs partitioning such that observations with similar characteristics fall in the same group. Let data at node m  be represented by $Q_m$. For each candidate split of the tree,$ j \in x_i$, where j is a possible threshold in feature $x_i$, the data is split into:
\begin{align}
    Q_m^{\text{left}} &= \{Q_m \, | \, x_i \leq j\} \\
    Q_m^{\text{right}} &= Q_m \setminus Q_m^{\text{left}}
\end{align}
and then quantile regression models are built for $Q^{left}_m$ and $Q^{right}_m$ respectively. The model chooses j as the threshold that minimizes the sum of the loss in the node using Equation(\ref{eqn:7}) across the possible values of j across the features of the dataset in the leaf nodes
\label{eqn:7}
\[
Loss =
\begin{cases}
  |r(y_i - \beta_0 - x_i^1 \beta_1)|, & \text{if } (y_i - \beta_0 - x_i^1 \beta_1) \leq 0 \tag{\ref{eqn:7}} \\
  |(1-r)(y_i - \beta_0 - x_i^1 \beta_1)|,   & \text{if } (y_i - \beta_0 - x_i^1 \beta_1) > 0  
\end{cases}
\]
where r is the quantile of estimation and $\beta_0$ and $\beta_1$ are estimates from the quantile regression.
The prediction of a new data point in this model is the quantile regression model prediction of the final leaf node in which the data point falls.
\subsection{The Algorithm}
The algorithm for building the tree is described below: Given the maximum depth of the tree has not been reached and the number of samples in a node $>$ 1

\begin{enumerate}
    \item Find the threshold at the node using the split criterion.
    \item Estimate the left and right quantile regression models.
    \item Calculate the loss using Equation \ref{eqn:7}
    \item Repeat step 1 for both left and right leaf nodes until maximum depth is reached or the number of samples = 1
\end{enumerate}

\begin{figure}
    \centering
    \includegraphics{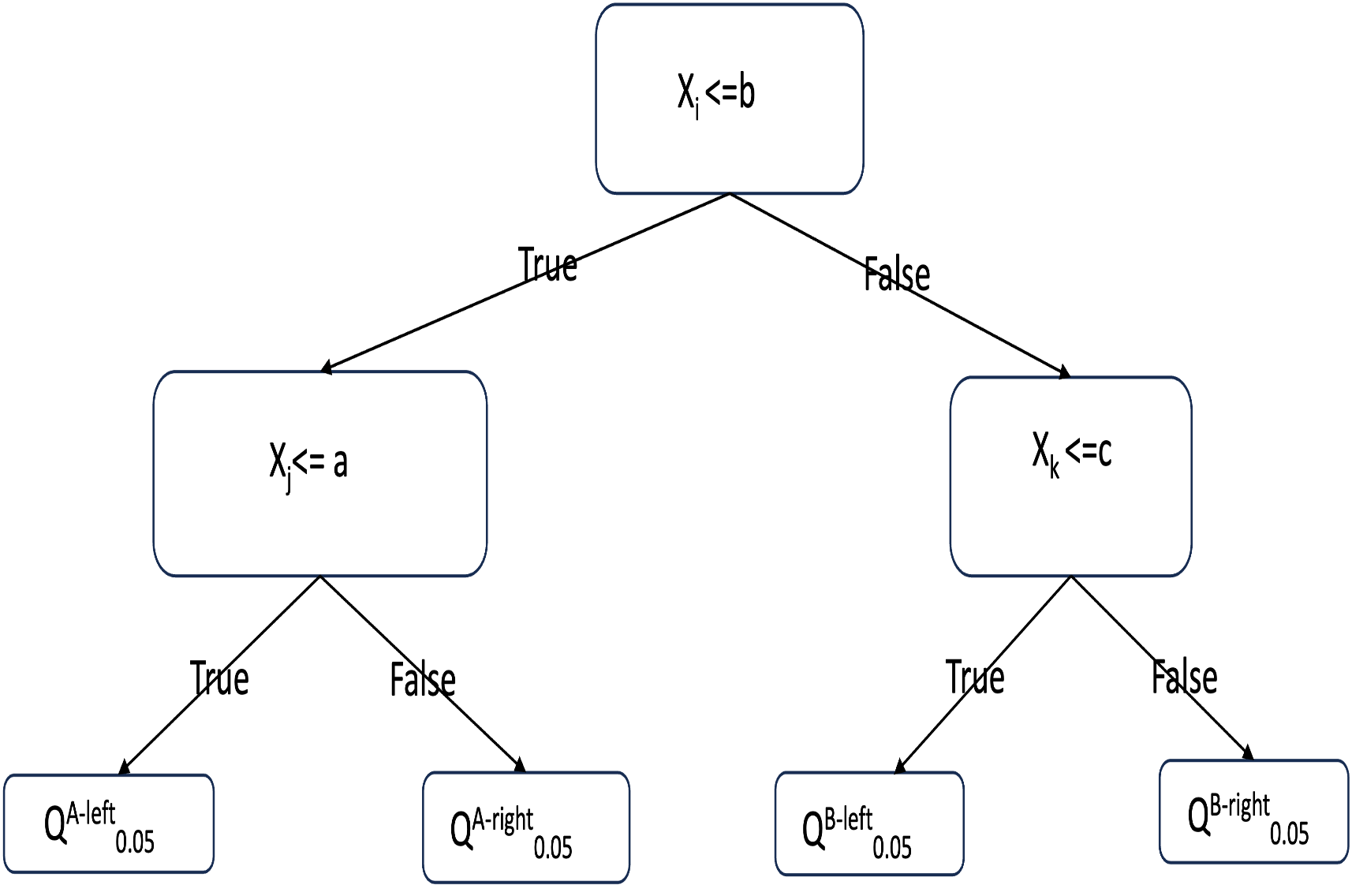}
    \caption{Sample of a quantile regression tree structure with depth = 2}
    \label{fig:model_sam}
\end{figure}

An example of a quantile regression tree structure for the $5^{th}$ quantile is shown in Figure \ref{fig:model_sam}

\section{Analysis and Results}
\subsection{Dataset}

Two datasets were used for metric evaluation of the model.

\begin{enumerate}
    \item Synthetic dataset: This dataset was generated to mimic a real-world scenario of  non-linear relationships among variables. The generator for the synthetic is:$ X \sim \mathcal{N}(0, 1)$, Y  = $(3X + e)^2$ and $e \sim \mathcal{N}(0, 1)$ where Y is the dependent variable of interest.
    \item Boston housing dataset: This dataset is gotten from kaggle\citep{boston_housing_dataset}. This comprises 506 observations and 14 variables with the dependent variable being the median value of owner-occupied homes in \$1000's and the feature variables including housing characteristics like per capita crime rate, proportion of owner-occupied units built before 1940, etc. After removing missing values, we are left we 501 observations and 14 variables.

\end{enumerate}


\subsection{Comparative Analysis}

To evaluate the effectiveness of the Quantileregressor tree (QRT), we compare its performance with several established models:

\begin{itemize}
    \item \textbf{Quantile Regression:} Provides a baseline for quantile estimation.  Directly comparing the QRT with a standard quantile regression model, emphasizes the benefits of the decision tree structure in the QRT.
    \item \textbf{Linear Regression:} Serves as a benchmark for average predictive performance. It provides a comparison to a traditional approach focusing on estimating the mean of the response variable.
    \item \textbf{Decision regressor tree :} Demonstrates the performance of a standard tree-based approach. It offers a comparison to a non-linear model without the quantile focus, highlighting the advantages of incorporating quantile regression in tree-based models.
\end{itemize}


\subsection{Performance Metrics}
\subsubsection{Using The Synthetic Dataset}
\begin{figure}[H]
  \centering
  \begin{subfigure}{0.45\linewidth}
    \centering
    \includegraphics[width=\linewidth]{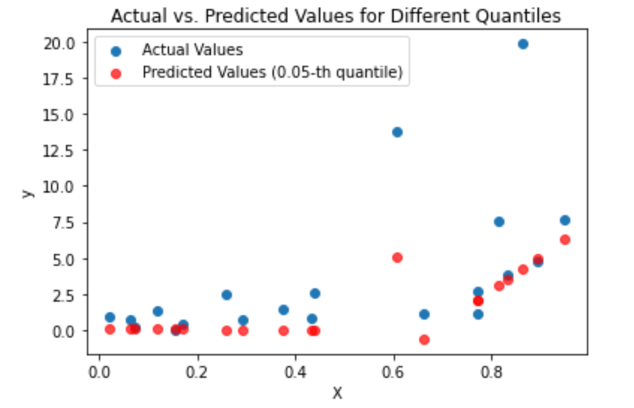}
    \caption{Quantile regression tree prediction at the 5th quantile}
  \end{subfigure}
  \hfill
  \hfill
  \begin{subfigure}{0.45\linewidth}
    \centering
    \includegraphics[width=\linewidth]{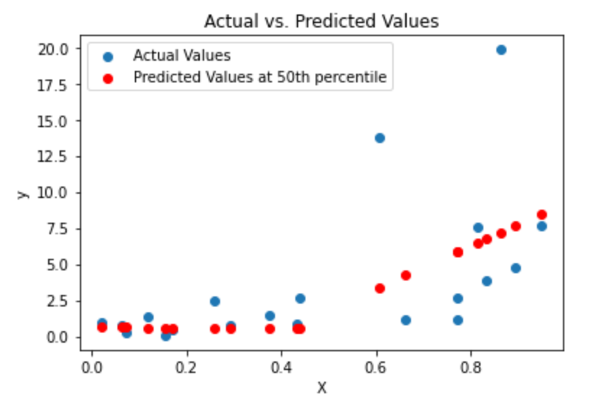}
    \caption{Quantile regression tree prediction at the 50th quantile}
  \end{subfigure}
  \hfill
  \hfill
  \begin{subfigure}{0.45\linewidth}
    \centering
    \includegraphics[width=\linewidth]{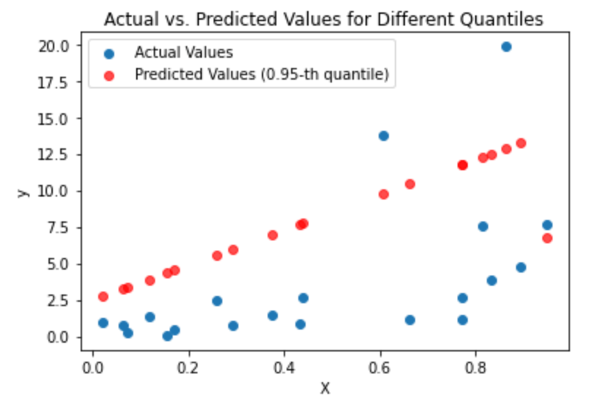}
    \caption{Quantile regression tree prediction at the 95th quantile }
  \end{subfigure}
  \caption{Quantile regression tree prediction at different quantiles}
  \label{fig:example}
\end{figure}

Figure \ref{fig:example} shows the predictions of the quantile regression tree model at different quantiles.

\begin{figure}[H]
    \centering
    \includegraphics[scale = 0.6]{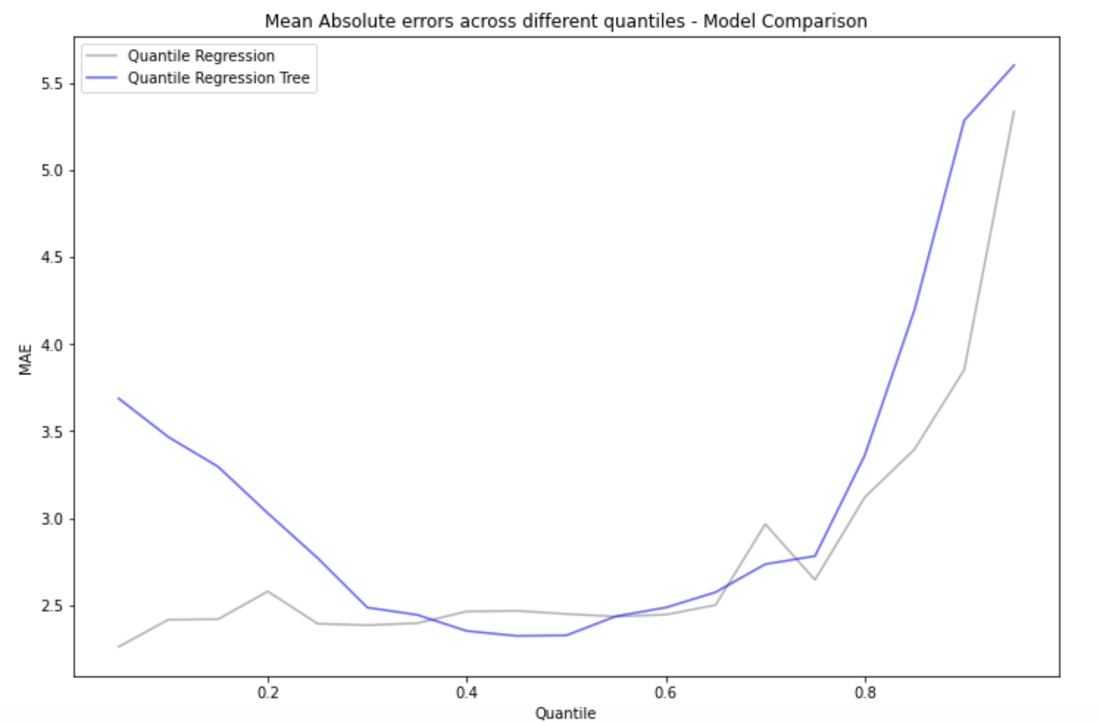}
    \caption{Mean absolute error comparison of quantile regression tree to quantile regression model using the synthetic data across different quantiles}
    \label{fig:qr_QRT_1}
\end{figure}

Figure \ref{fig:qr_QRT_1} shows a comparison of quantile regression tree to quantile regression using mean absolute error of the predictions on the synthetic dataset across different quantiles. The figure shows that on the synthetic data, the two models have fairly the same performance.

Further comparing the quantile estimate at the 50th quantile for quantile regression tree to the estimation of the linear regression model and decision regression tree, we have Table \ref{tab:performance_metrics}. Table \ref{tab:performance_metrics} uses mean absolute error (MAE) and mean squared error (MSE) as the primary metrics for assessing model performance. 

\begin{table}[h]
\centering
\caption{Comparative Performance Metrics}
\label{tab:performance_metrics}
\begin{tabular}{lcc}
\hline
Model               & Mean Absolute Error (MAE) & Mean Squared Error (MSE) \\ \hline
Quantile Tree       & 2.45                     & 17                    \\
Quantile Regressor   & 2.3                     & 19                    \\
Linear Regression      & 2.5                     & 17                   \\
Decision Tree  & 2.7                     & 23                    \\ \hline
\end{tabular}
\end{table}

\begin{itemize}
    \item \textbf{Quantile Tree vs. Quantile Regressor:} The traditional Quantile Regressor exhibits the lower MAE but the quantile regression tree performs better when using MSE. 
    
    \item \textbf{Quantile Tree vs. Linear Regression:} The QRT performs better when using MAE but the linear regression model and the QRT have similar performance when using  MSE.
    
    \item \textbf{Quantile Tree vs. Decision Tree:} The QRT outperforms the Decision Tree in both MAE and MSE, indicating its superior performance in capturing the conditional distribution of the dependent variable, especially in more complex data scenarios.

\end{itemize}

\subsubsection{Using Boston Housing Dataset}
\begin{figure}[H]
  \centering
  \begin{subfigure}{0.8\linewidth}
    \centering
    \includegraphics[width=\linewidth]{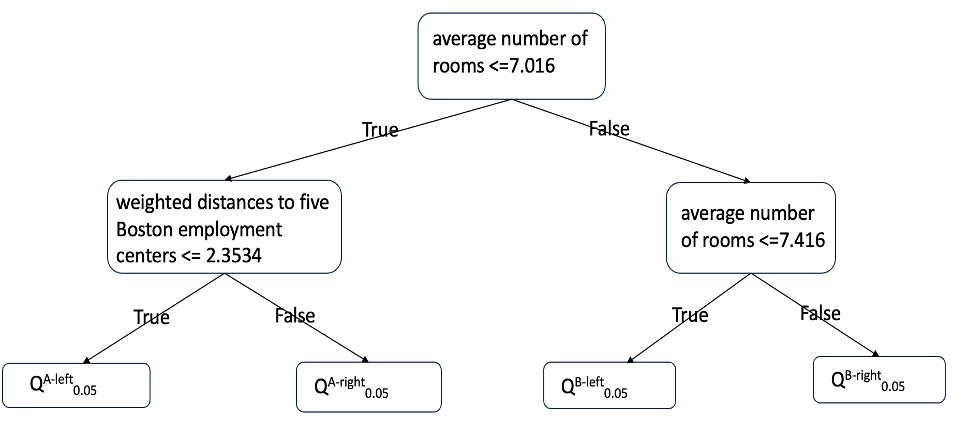}
    \caption{Tree structures of QRT at the 5th quantile}
  \end{subfigure}
  \hfill
  \begin{subfigure}{0.8\linewidth}
    \centering
    \includegraphics[width=\linewidth]{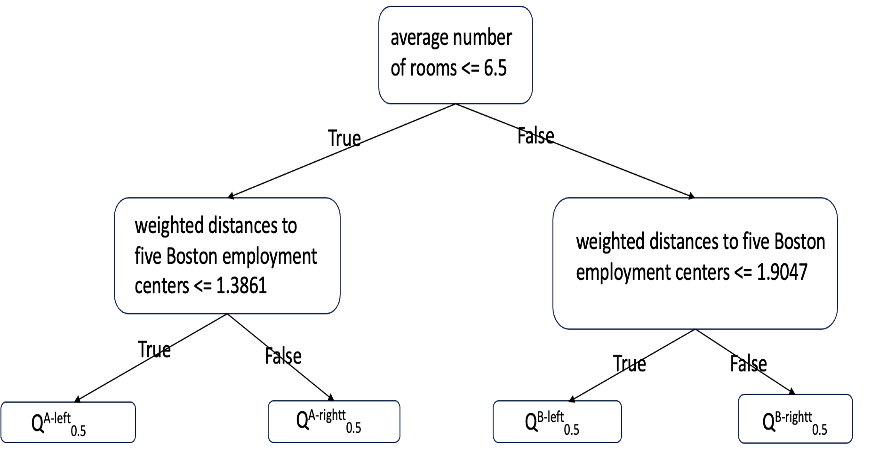}
    \caption{Tree structures of QRT at the 50th quantile}
  \end{subfigure}
  \hfill
  \begin{subfigure}{0.8\linewidth}
    \centering
    \includegraphics[width=\linewidth]{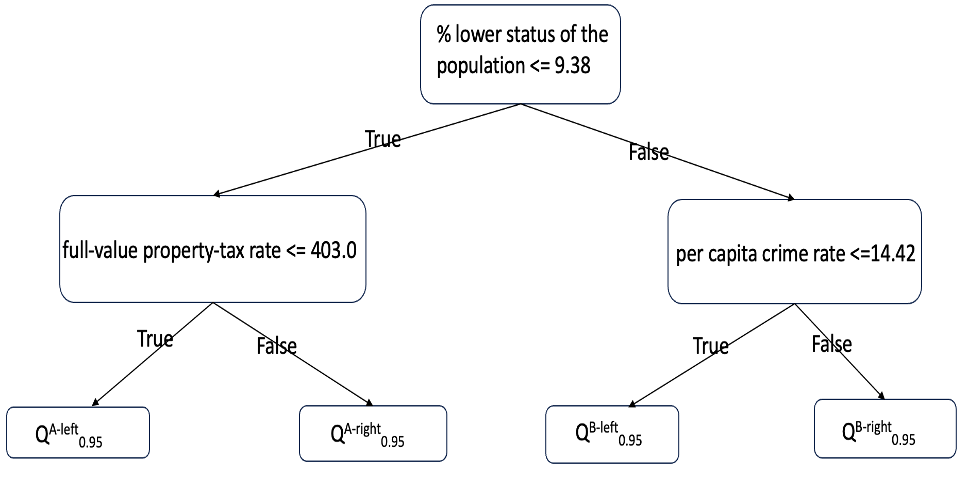}
    \caption{Tree structures of QRT at the 95th quantile}
  \end{subfigure}
  \caption{Tree structures of QRT at different quantiles}
  \label{fig:tree_1}
\end{figure}

Figure \ref{fig:tree_1} shows the different tree structures for the quantile regression tree for estimating different quantiles. The differences in tree structure suggest that the variables that best estimate the quantiles differ as quantiles differ. This suggests the presence of a non-linear relationship in the dataset when predicting quantiles.

\begin{figure}[H]
    \centering
    \includegraphics[scale =0.6]{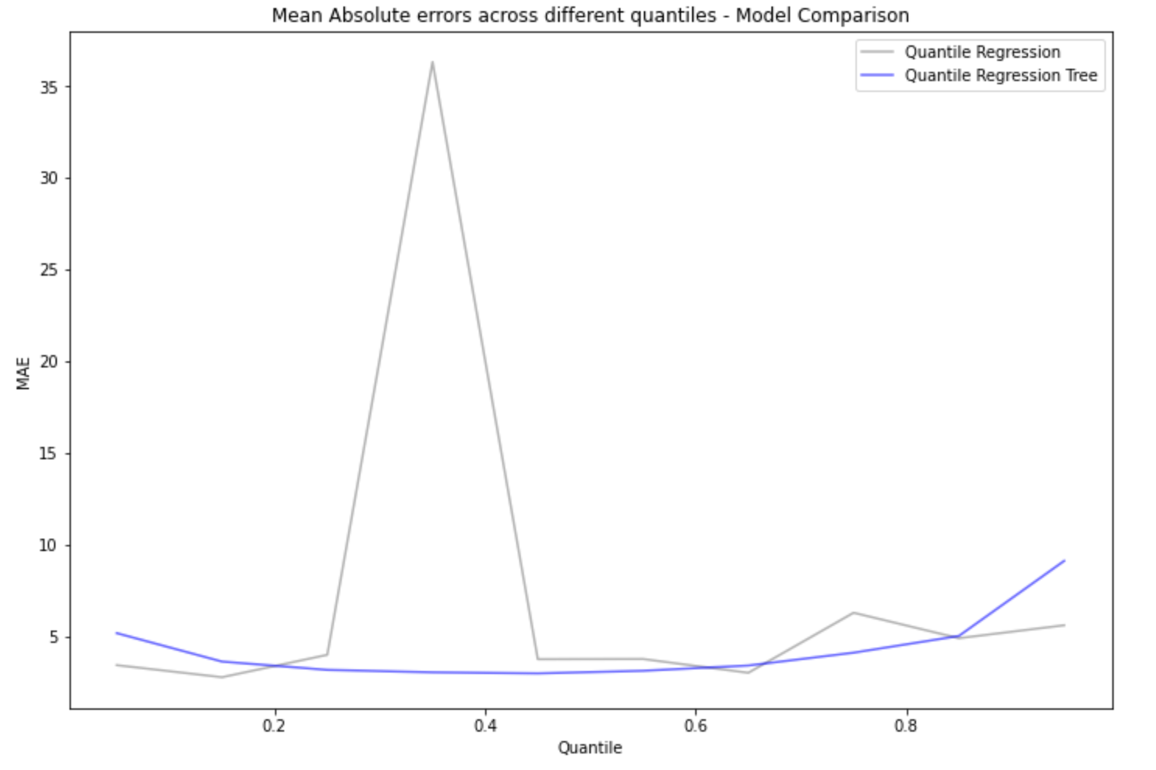}
    \caption{Caption}
    \label{fig:qr_QRT_2}
\end{figure}

Figure \ref{fig:qr_QRT_2} shows a comparison of QRT to quantile regression using mean absolute error of the predictions on the Boston housing dataset across different quantiles. The figure shows that on the Boston housing dataset, the QRT slightly outperforms the quantile regression model, especially around the 40th quantile.

\begin{figure}[H]
  \centering
  \begin{subfigure}{0.6\linewidth}
    \centering
    \includegraphics[width=\linewidth]{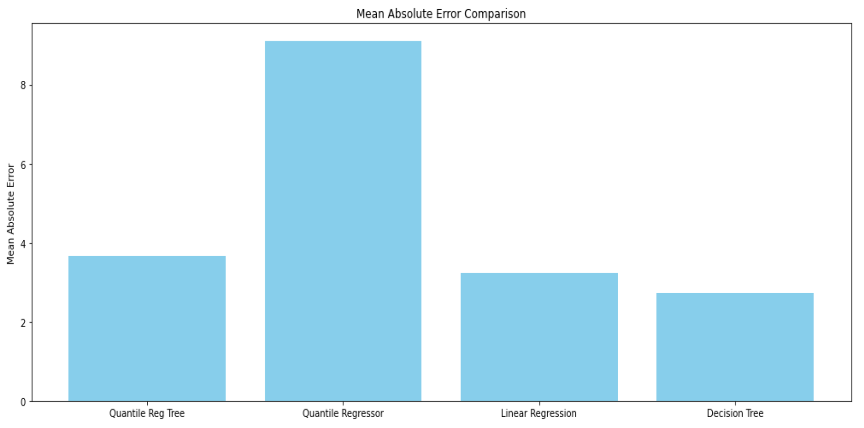}
    \caption{Mean absolute error comparison across models}
  \end{subfigure}
  \hfill
  \begin{subfigure}{0.6\linewidth}
    \centering
    \includegraphics[width=\linewidth]{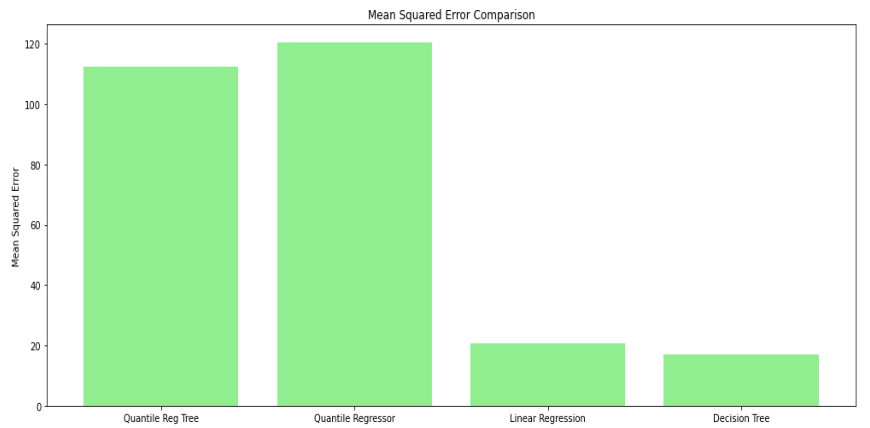}
    \caption{Mean sqaured error comparison across models}
  \end{subfigure}
  \caption{Tree structures of QRT at different quantiles}
  \label{fig:mae_mse}
\end{figure}

Figure \ref{fig:mae_mse} shows that the QRT outperforms the quantile regression model when considering both MSE and MAE's. But, for this dataset, the linear regression model and random forest models outperform the QRT.

Generally, the quantile regressor tree (QRT) demonstrates comparative effectiveness in handling non-linear data structures, a conclusion supported by its performance when compared to the traditional models. However, in some scenarios, the traditional models may yield better results. Consequently, the selection of an appropriate model should be guided by the specific characteristics of the dataset and the research question at hand, with particular consideration given to whether the focus is on specific quantiles or the overall distribution.

\section{Conclusion}
This study introduced and assessed the quantile regressor tree (QRT), an innovative approach that combines the robustness of quantile regression with the interpretability and adaptability of decision trees. Through empirical analysis, we compared the QRT's performance with that of conventional regression models, such as quantile regression, linear regression, and decision regressor tree. We conducted this evaluation using synthetic datasets designed to simulate non-linear relationship scenarios, as well as the well-known Boston housing dataset.

The QRT demonstrated its effectiveness in capturing the conditional distribution of the response variable, particularly excelling in complex or non-linear scenarios. Despite displaying higher Mean Absolute Error (MAE) and Mean Squared Error (MSE) in certain conditions compared to some traditional models, its notable strength lies in its ability to focus on specific quantiles. This particular capability renders it particularly valuable in applications where understanding the distribution tails or specific quantile behaviors is of paramount importance.

\bibliographystyle{plainnat}
\bibliography{sample}

\end{document}